\documentclass{jps-cp}
\usepackage{txfonts} %Please comment out this line unless the txfonts package is availabe in your LaTeX system.

\title{Irrelevance of Anomalous Breaking of Axial U(1) Symmetry and the U(1) Problem}

\author{Nodoka \textsc{Yamanaka}$^{1,2}$}

\inst{$^{1}$Department of Physics, Graduate School of Science, Tohoku University, Sendai 980-8578, Japan \\
$^{2}$Nishina Center for Accelerator-Based Science, RIKEN, Wako 351-0198, Japan}

\email{nodoka.yamanaka@riken.jp}

\recdate{January 17, 2026}

\abst{
The $\eta$ and $\eta'$ mesons are conventionally known to receive contribution from the anomalous breaking of axial $U(1)$ symmetry, and they are considered to not be the Nambu-Goldstone (NG) bosons of the spontaneous chiral $SU(3)_L \times SU(3)_R$ symmetry breaking of QCD. 
However, it has recently been shown that this axial $U(1)$ anomaly is not actually physical. 
In this contribution, we first review this statement and then propose a mechanism in which $\eta$ and $\eta'$ mesons are indeed NG bosons while being consistent with the axial U(1) problem.
}

\kword{Chiral anomaly, U(1) problem, $\eta$ meson, $\eta'$ meson}

\begin{document}
\maketitle

\section{Introduction}

Hadron masses are believed to be generated by the dynamics of quantum chromodynamics (QCD), the fundamental theory of the strong interaction, and numerical simulations are in good agreement with experimental data \cite{PACS-CS:2008bkb}.
It is still difficult to quantify phenomena due to the nonperturbative effects of QCD, but we know that the spontaneous chiral symmetry breaking \cite{Nambu:1961tp,Nambu:1961fr,Goldstone:1961eq,Goldstone:1962es} and the chiral perturbation \cite{Gasser:1983yg,Gasser:1984gg} relying on small current quark masses can be used to qualitatively explain the hadron spectrum.
Particularly, light pseudoscalar mesons are identified as the Nambu-Goldstone (NG) modes.
However, $\eta'$, despite its quantum number of axial $U(1)$ NG boson, is much heavier than the other light ones, while another isoscalar meson, $\eta$, is not as light as the pion, and they have been the central figures of the long-standing ``U(1) problem'' \cite{Weinberg:1975ui,Christos:1984tu}.
The most popular resolution is the axial $U(1)$ anomaly \cite{Bardeen:1969md} and the instanton \cite{tHooft:1976rip}, but it has recently been pointed out that this anomalous breaking of axial $U(1)$ symmetry is actually irrelevant \cite{Yamanaka:2022vdt,Yamanaka:2022bfj}.
In this contribution, we propose an alternative scenario where the U(1) problem is resolved without axial anomaly, and state that the $\eta'$ meson (and also $\eta$) is actually an NG boson \cite{Yamanaka:2024nzn}.
In the next section, we review the spontaneous chiral symmetry breaking and the mechanism generating the masses of the NG bosons in QCD.
We then introduce the details of the U(1) problem in Sec. \ref{sec:U(1)problem}.
In Sec. \ref{sec:resolution}, we show how $\eta$ and $\eta'$ acquire their masses by avoiding the conventional U(1) problem.
The final section is devoted to the summary.

\section{Spontaneous Chiral Symmetry Breaking}

\begin{figure}[tbh]
\begin{center}
\includegraphics[width=12cm]{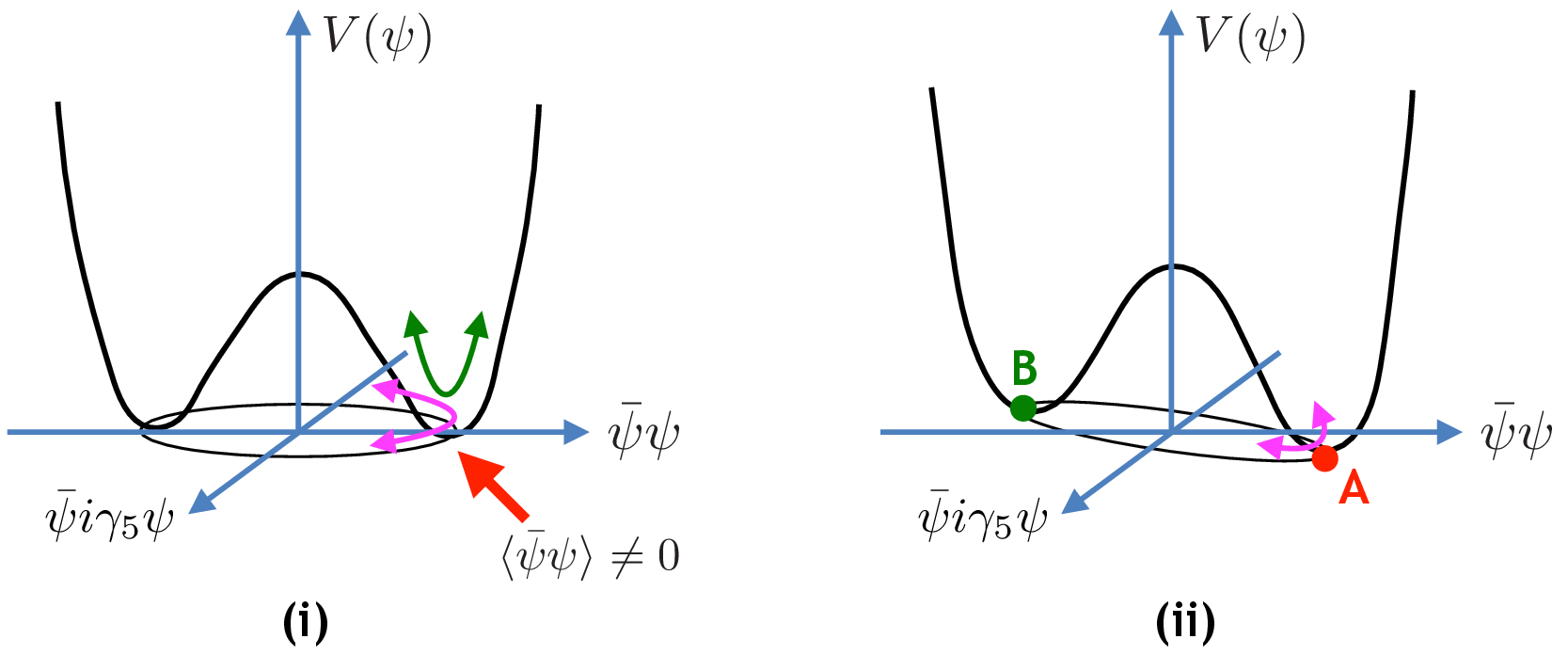}
\caption{
QCD effective potential in the $(\bar \psi \psi , \bar \psi i \gamma_5 \psi )$ space.
The case for massless quarks is shown on the left (i), and the case where the QCD Lagrangian has a finite current quark mass on the right (ii).
The point $(\bar \psi \psi , \bar \psi i \gamma_5 \psi )=(0,0)$ is unstable, and the true vacuum is at the bottom of the valley.
The radial direction has a large positive curvature (green double arrow of left panel) which indicates a large dynamical quark mass.
The circular direction is flat (pink double arrow of left panel) which corresponds to the massless NG mode.
For the massive current quark case (right panel), the chiral circle is tilted and there is only one true vacuum (point A), while the opposite side has an unstable saddle point (point B).
At the point A, there is a small positive curvature in the circular direction, which corresponds to the small (pseudo-)NG boson mass.
}
\label{fig:QCD_potential_2}
\end{center}
\end{figure}

The spontaneous chiral symmetry breaking of QCD is qualitatively explained by the strong attraction between quarks and antiquarks which renders the vacuum unstable.
When the binding energy between the quark $q$ and antiquark $\bar q$ exceeds their masses, the perturbative vacuum becomes energetically favorable for generating more and more $q \bar q$ pairs.
This fact is translated to an effective negative coefficient of the operator $(\bar q q )^2$ of the QCD effective potential.
When the number density of $q \bar q$ pairs increases, the effect of repulsive higher power interactions $(\bar q q )^n$ $(n>2)$, for instance due to the Pauli exclusion, becomes important so that the growth of $\bar q q$ eventually stops at an expectation value $\langle \bar q q \rangle$.
We also note that QCD has chiral symmetry, so we should have a rotationally symmetric potential in the $(  \bar q q , \bar q i\gamma_5 q )$ space.
The qualitative shape of the QCD effective potential (the so-called ``Mexican hat'' or ``wine bottle'' potential) is shown in Fig. \ref{fig:QCD_potential_2} (i).
Now we have a new vacuum where $\langle \bar q q \rangle$ is finite (we should properly write $\langle \cos \alpha \,  \bar q q + \sin \alpha \, \bar q i\gamma_5 q \rangle$, but we can always redefine the arbitrary angle $\alpha$ to zero thanks to the chiral symmetry), which is the spontaneous chiral symmetry breaking.
The QCD potential at the true vacuum point has a positive curvature in the radial direction, which is interpreted as the dynamical (or ``constituent'') quark mass.
Another remarkable point is that the circular direction is flat due to the chiral symmetry (also called ``chiral circle''), and the vacuum is infinitely degenerate.
This actually implies that a massless mode exists, which is just the NG boson.

\begin{figure}[tbh]
\begin{center}
\includegraphics[width=8cm]{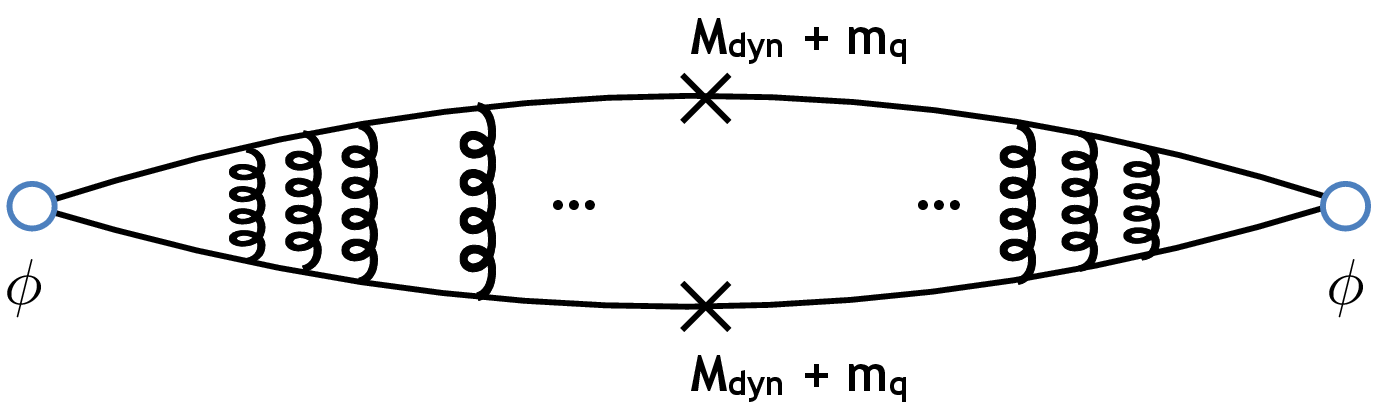}
\caption{
Schematic picture of the (connected) meson correlator.
The solid and wiggly lines are the quark and gluon propagators, respectively.
It is assumed that there are infinite gluon exchanges.
The blobs and the crosses are the interpolating operators of the meson $\phi$ and the quark mass insertions, respectively.
}
\label{fig:meson_correlator}
\end{center}
\end{figure}

The QCD Lagrangian has small current quark masses, so the chiral symmetry is slightly broken.
Since the quark mass term $-m_q \bar q q$ is linear in $\bar qq$, the chiral circle of the QCD effective potential becomes tilted [see Fig. \ref{fig:QCD_potential_2} (ii)], so that the vacuum is no longer degenerate and the NG boson acquires a small mass.
The mass of the NG mode may also diagrammatically be explained.
The meson correlator (or propagator) is essentially a quark loop where gluons are exchanged (see Fig. \ref{fig:meson_correlator}).
For the case of the NG boson, the dynamical mass of the quark and antiquark is exactly canceled by the binding energy due to the gluonic attraction, which makes the system massless.
However, when the current quark mass is finite, there is an additional mass insertion which is not canceled.
Since the dynamical quark mass $M_{\rm dyn}$ is much larger than the current quark one $m_q$, the meson mass squared is approximately proportional to $M_{\rm dyn} m_q$ which explains the empirically known mass dependence $\propto \sqrt{m_q}$ of the (pseudo-)NG boson.

\section{The U(1) Problem\label{sec:U(1)problem}}

Within the leading order chiral perturbation, i.e. the linear approximation of current quark mass, it is possible to calculate the physical mass spectrum of the NG bosons.
In the $SU(3)_L \times SU(3)_R$ basis, the charge neutral mesons have three components, namely $\pi_0 , \eta_8$, $\eta_0$.
Their mass matrix is given by
\begin{eqnarray}
&&
\hspace{-2em}
M_{\rm conn}^2
=
\left(
\begin{array}{ccc}
2 m_l B_{\rm conn} & 0 & 0 \cr
0 & \frac{2}{3} (m_l + 2 m_s) B_{88} & \frac{2\sqrt{2}}{3} (m_l - m_s) B_{80} \cr
0 & \frac{2\sqrt{2}}{3} (m_l - m_s) B_{80} & \frac{2}{3} (2m_l + m_s) B_{00} \cr
\end{array}
\right)
,
\label{eq:NGmassmatrix1}
\end{eqnarray}
where $B_{\rm conn} =2.6$ GeV is a low energy constant which was fitted from the squared pion mass $m_\pi^2 = 2 m_l B_{\rm conn} = (135\, {\rm MeV})^2$.
The other coefficients are given by $B_{ij} = \frac{f_\pi^2}{f_i f_j} B_{\rm conn}$ $(i,j=0,8)$, where $f_\pi = 92.2$ MeV, $f_8 = 115$ MeV, and $f_0 = 100$ MeV \cite{Bali:2021qem}.
Here we used $m_s = 93.5$ MeV, while $m_l \equiv \frac{1}{2} (m_u +m_d)= 3.43$ MeV is the averaged up and down quark masses, and we neglected the mass splitting $m_d -m_u$.
In addition to the already diagonalized pion mass (135 MeV), we expect to obtain the mass of $\eta$ ($m_\eta = 550$ MeV) and $\eta'$ ($m_{\eta'} = 960$ MeV) after the diagonalization of the above matrix, but we only get in the actual calculation one heavy ($\approx 600$ MeV) and another light mass (120 MeV), which clearly do not agree with the experimental data.
The light meson generated from $\eta_8 - \eta_0$ mixing can be explained as follows.
If we neglect the up and down quark masses and set $f_\pi = f_8 =f_0$, the mass matrix (\ref{eq:NGmassmatrix1}) is proportional to
\begin{eqnarray}
M_{\rm conn}^2
&\propto &
\left(
\begin{array}{ccc}
0 & 0& 0 \cr
0 & 2 & -\sqrt{2} \cr
0 & -\sqrt{2} & 1 \cr
\end{array}
\right)
,
\label{eq:NGmassmatrix1.5}
\end{eqnarray}
for which the last two lines are linearly dependent.
This means that, in addition to the first line which trivially has a zero eigenvalue, there is another degenerate eigenmode having a vanishing eigenvalue.
These two massless modes yield two light mesons with mass 135 MeV or below if we retain $m_l$ finite and $f_\pi \neq f_8 \neq f_0$.
This absence of ``not light'' $\eta$ ($m_\eta = 550$ MeV) is just the U(1) problem [in the literature, the existence of heavy $\eta'$ ($m_{\eta'} = 960$ MeV) is often highlighted as the problem].
We also note that this is stated in a different way in the original work of Weinberg \cite{Weinberg:1975ui} (see Appendix for details).

To resolve the U(1) problem, we need to break the linear dependence of the last two lines of the mass matrix (\ref{eq:NGmassmatrix1.5}).
For this, adding the contribution from the topological susceptibility \cite{Witten:1979vv,Veneziano:1979ec,Veneziano:1980xs}
\begin{eqnarray}
U(k)
&=&
i\frac{3 \alpha_s^2}{32\pi^2 f_\pi^2} 
\int d^4 x\, e^{ik\cdot x} \langle 0| 
F_{\mu \nu, a}\tilde F^{\mu \nu}_a (x) 
F_{\rho \sigma, b}\tilde F^{\rho \sigma}_b (0) 
|0\rangle
,
\label{eq:topological_susceptibility}
\end{eqnarray}
where $F^{\mu \nu}_a$ and $\tilde F^{\mu \nu}_a$ are the gluon field strength and its dual, with $\alpha_s \equiv \frac{g_s^2}{4\pi }$ the strong coupling, and the Kobayashi-Maskawa-'t Hooft interaction \cite{tHooft:1976rip,Kobayashi:1970ji,Kobayashi:1971qz} 
\begin{equation}
{\cal L}_{\rm KMT}
\propto
\bar u_R u_L \bar d_R d_L \bar s_R s_L + {\rm h.c.}
,
\label{eq:KMT}
\end{equation}
was proposed.
Since both effects are due to the chiral anomaly, the nonconservation of the flavor singlet axial charge \cite{Bardeen:1969md}
\begin{equation}
\sum_{q=u,d,s}
\Bigl[
\partial^\mu (\bar q \gamma_\mu \gamma_5 q )
+2m_q
\bar q i\gamma_5 q
\Bigr]
=
-\frac{3 \alpha_s}{8\pi}F_{\mu \nu, a}\tilde F^{\mu \nu}_a
,
\label{eq:chiralWTI}
\end{equation}
the mass matrix (\ref{eq:NGmassmatrix1}) will receive an extra term only in the $\eta_0$ channel, of the form $M^2_{\rm anom}=$ diag[0,0,$U$], which can break the linear dependence of the second and third lines.
This shift is conventionally justified by claiming that the axial $U(1)$ symmetry is explicitly broken so that $\eta_0$ is not anymore an NG boson.
The diagonalization of $M_{\rm conn}^2+M^2_{\rm anom}$ with the fitted parameter $U= (860\, {\rm MeV})^2$ yields $m_{\eta} = 420$ MeV and  $m_{\eta'} = 960$ MeV which are in reasonable agreement with experimental data.

However, as we mentioned in the Introduction, the chiral anomaly cannot break the global axial $U(1)$ symmetry \cite{Yamanaka:2022vdt,Yamanaka:2022bfj,Yamanaka:2024nzn}.
It was believed so far that this explicit violation is due to the nontrivial topological defects of the gauge configuration which are probed with the following topological charge
\begin{eqnarray}
\frac{\alpha_s}{8\pi}
\int d^4 x\,
F_{\mu \nu, a}\tilde F^{\mu \nu}_a
&=&
\frac{i g_s \alpha_s}{24\pi} 
\int d^3 \vec{x}\, 
f_{abc} \epsilon_{ijk} A_{ia} (\vec{x}) A_{jb} (\vec{x}) A_{kc} (\vec{x})
\bigg|_{t=-\infty}^{t=+\infty} 
=
\Delta n
,
\label{eq:topologicalcharge}
\end{eqnarray}
which is always an integer number.
We remark that the second equality is a triple product of gauge field components $A_a^\mu$ in the 3-dimensional space, so it always contains the unphysical longitudinal gauge mode (proportional to the momentum), so Eq. (\ref{eq:topologicalcharge}) is not an observable quantity (see Refs. \cite{Yamanaka:2022vdt,Yamanaka:2022bfj} for more rigorous derivations).
Since the anomalous violation of axial $U(1)$ symmetry (\ref{eq:chiralWTI}) is due to the topological charge, this explicit breaking is irrelevant, and we therefore need other mechanisms to resolve the U(1) problem.

\section{Our solution to the U(1) Problem\label{sec:resolution}}

\begin{figure}[tbh]
\begin{center}
\includegraphics[width=10cm]{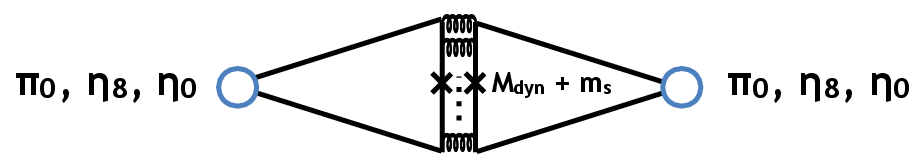}
\caption{
Schematic picture of the disconnected diagram made of two quark loops.
In the middle, there is a quark-antiquark pair propagating in the t-channel with mass insertions and all possible gluon exchanges.
}
\label{fig:meson_correlator_disc}
\end{center}
\end{figure}

We propose to resolve the U(1) problem by considering the disconnected diagram as shown in Fig. \ref{fig:meson_correlator_disc}.
This process has two quark loops and a quark-antiquark pair propagates in the t-channel, where we again have the exchange of gluons and mass insertions in the quark lines.
In the leading order of chiral perturbation, the disconnected contribution to the NG boson mass matrix is \cite{DiVecchia:1980vpx,Christos:1984tu}
\begin{eqnarray}
&&
M_{\rm disc}^2
=
\left(
\begin{array}{ccc}
0 & 0& 0 \cr
0 & 0& \sqrt{2} (m_l -m_s) B'_{80} \cr
0 & \sqrt{2}(m_l -m_s) B'_{80} & 2 (2m_l+m_s) B'_{00} \cr
\end{array}
\right)
,
\label{eq:NGmassmatrix3}
\end{eqnarray}
where $B'_{ij} \equiv \frac{f_\pi^2}{f_i f_j} B_{\rm disc}$ $(i,j=0,8)$ are low energy constants.
The second and third lines of this matrix are linearly independent even if $f_8=f_0$, so it is also a candidate for resolving the U(1) problem.
We then obtain nondegenerate masses of the pion, $\eta$, and $\eta'$ mesons by summing with the connected part (\ref{eq:NGmassmatrix1}) and diagonalizing the matrix.
We fit $B_{\rm disc} = -5.6$ GeV so as to obtain $m_{\eta'} = 960$ MeV.
We then obtain $m_\eta^2 = -(570\, {\rm MeV})^2$ for the mass of $\eta$, which is in good agreement with the observed value except the sign.
Naively the negative squared mass seems to be pathologic, but this actually corresponds to the negative curvature of the saddle point of the QCD effective potential [see point B of the Fig. \ref{fig:QCD_potential_2} (ii)].
Since the tilt of the chiral circle is small, we should have another true vacuum in the opposite side of the $(\bar qq , \bar q i\gamma_5 q)$ space which has a positive curvature with almost the same magnitude [point A of Fig. \ref{fig:QCD_potential_2} (ii)].
At this true vacuum, the mass of $\eta$ is $m_\eta^2 = +(570\, {\rm MeV})^2$, which resolves the U(1) problem.
The mass matrix (\ref{eq:NGmassmatrix3}) is part of the chiral Lagrangian 
\begin{equation}
{\cal L}_{\rm disc}
=
\frac{i\sqrt{3}}{2\sqrt{2}}
B_{\rm disc} f_\pi
\eta_0 
{\rm Tr} 
\Biggl[ 
M_q 
\exp \Biggl(
\sum_{a=0}^{8}
\frac{i\lambda_a \eta_a}{f_\pi} 
\Biggr)
\Biggr]
+{\rm h.c.}
.
\label{eq:chiral_lagrangian}
\end{equation}
which was previously analyzed as a higher order interaction of the $1/N_c$ expansion \cite{DiVecchia:1980vpx,Christos:1984tu}.
Interestingly, the disconnected contribution (\ref{eq:NGmassmatrix3}) is numerically dominant with $B_{\rm disc} = -5.6$ GeV, so the large $N_c$ argument is not effective at the leading order.

To demonstrate the physical viability of our mechanism, we calculate the two-photon decay rates of the $\eta$ and $\eta'$ mesons using the $\eta -\eta'$ mixing angle of approximately -18$^\circ$ obtained by diagonalizing $M_{\rm conn}^2+M_{\rm disc}^2$ and compare with experimental data.
The results, shown in Table \ref{table:decay}, are in good agreement.

\begin{table}[tbh]
\caption{
Comparison of the two-photon decay rates obtained in our work ($\theta = -18^\circ$) and the experimental values.
We also show the predictions admitting the chiral anomaly seen in Sec. \ref{sec:U(1)problem} ($\theta = -13^\circ$).
The unit is in eV.
}
\begin{center}
\begin{tabular}{l|ccc}
$\Gamma$ & Chiral anomaly ($\theta = -13^\circ$) & 
Our work ($\theta = -18^\circ$) & Experiments \cite{ParticleDataGroup:2024cfk}  \\
\hline
$\eta \to \gamma \gamma $ & 320 & 418 & 516(20) \\
$\eta' \to \gamma \gamma $ & 5170 & 4640 & 4340(140) \\
\end{tabular}
\end{center}
\label{table:decay}
\end{table}

\section{Summary}

In this contribution, we discussed the resolution of the U(1) problem.
This problem is due to the fact that the connected meson correlator has an almost linear dependence in the $\eta - \eta'$ mixing sector, so that the physical $\eta$ has a too small mass eigenvalue.
Conventionally, the large $\eta$ (550 MeV) and $\eta'$ (960 MeV) masses are explained by the chiral anomaly.
However, the present author has recently shown that the topological effect of QCD is not observable, which implies that the explicit axial $U(1)$ symmetry breaking by the chiral anomaly does not occur, and another mechanism to explain the masses of $\eta$ and $\eta'$ was needed.
For this, we considered the disconnected diagram of the meson correlator which breaks the linear dependence of the $\eta - \eta'$ mixing sector.
We finally found $m_\eta = 570$ MeV and $m_{\eta'} = 960$ MeV, in good agreement with the experimental data.
In the intermediate step of the derivation, we also encountered a negative squared mass which corresponds to the saddle point of the QCD effective potential, but the sign could have just been inverted thanks to the approximate chiral symmetry.
We also calculated the two-photon decay of $\eta$ and $\eta'$ which agreed with the experiments as well.
In future studies, we may use the fitted chiral Lagrangian without the axial anomaly to predict observables related to these mesons such as the three-pion decay \cite{Gan:2020aco} or mesic nuclei \cite{Itahashi:2012ut,Nagahiro:2012aq,Metag:2017yuh}.
Another interesting test is the 2-flavor lattice QCD calculation which is less costly than the 2+1-flavor QCD so that the light quark masses may be varied more easily.

This work (project) was supported by the RIKEN TRIP initiative (Nuclear transmutation).

\appendix
\section{Weinberg's formulation of the U(1) problem\label{sec:WeinbergU(1)problem}}

We briefly summarize the derivation of the U(1) problem by Weinberg \cite{Weinberg:1975ui}.
Given $m_L^2$ the lowest squared mass eigenvalue of the connected mass matrix (\ref{eq:NGmassmatrix1}), the following equality holds for an arbitrary normalized state vector $| \psi \rangle$:
\begin{equation}
\langle \psi | M_{\rm conn}^2 | \psi \rangle
\ge m_L^2
.
\end{equation}
At low energy, a state vector is given by the decay constant which is the wave function at the origin.
By substituting the isovector state vector $| \phi \rangle =( 0, f_8 , \sqrt{2} f_0) / \sqrt{3}$ [note that $\frac{1}{\sqrt{2}}| \bar u u + \bar d d \rangle =( 0, 1 , \sqrt{2}) / \sqrt{3}$ in the $SU(3)$ nonet basis], the inequality becomes
\begin{equation}
\langle \phi | M_{\rm conn}^2 | \phi \rangle
= 
2 m_l f_\pi^2 B_{\rm conn}
=
f_\pi^2 m_\pi^2
\ge 
m_L^2 \langle \phi | \phi \rangle
=
\frac{1}{3}
(f_8^2+2f_0^2) m_L^2
.
\end{equation}
In the extreme case where $f_0 =0$ and $f_8 = f_\pi$, we have 
\begin{equation}
3 m_\pi^2
\ge m_L^2
.
\label{eq:WeinbergU(1)problem}
\end{equation}
If we assign $m_L = m_\eta$, there is clearly a problem: this is the U(1) problem of Weinberg.
An important assumption of this derivation is the positivity of the squared mass eigenvalues.
The discussion of this contribution actually resolves the U(1) problem by allowing negative eigenvalues, which is still consistent with the constraint (\ref{eq:WeinbergU(1)problem}).

\end{document}